\documentclass[twocolumn,prl]{revtex4}

\newcommand{\beq}{\begin{equation}}
\newcommand{\eeq}{\end{equation}}
\newcommand{\beqa}{\begin{eqnarray}}
\newcommand{\eeqa}{\end{eqnarray}}

\def\half{\frac{1}{2}}
\def\<{\langle}
\def\>{\rangle}

\def\opone{\leavevmode\hbox{\small1\kern-3.8pt\normalsize1}}

\def\Chi{{\chi}}
\def\ket#1{|#1\rangle}

\begin{document}
\title{Quantum gloves: Physics and Information}
%=================================================================

\author{N. Gisin\\
\protect\small\em Group of Applied Physics, University of Geneva,
1211 Geneva 4, Switzerland}
\date{\today}

\begin{abstract}
The slogan {\it information is physical} has been so successful
that it led to some excess. Classical and quantum information can
be thought of independently of any physical implementation. Pure
information tasks can be realized using such abstract c- and
qu-bits, but physical tasks require appropriate physical
realizations of c- or qu-bits. As illustration we consider the
problem of communicating chirality.
\end{abstract}

\maketitle

%\begin{center}
%{\tt PACS 03.65.Bz, 89.70.+c, 42.50.Wm}
%\end{center}
%\begin{multicols}{2}

%\section{Introduction}\label{int}
%================================
Assume that two distance partners like to compare the chiralities
of their Cartesian reference frames. This is impossible by
exchanging only classical information, i.e. by sending only
abstract 0's and 1's. This is true in the most general
relativistic context \cite{Feynman}. But for our purpose it
suffice to consider a Newtonian physics worldview. It is then
quite intuitive to grasp why information about chirality can't be
encoded in classical bits: bits measure the quantity of
information, but have per se no meaning, in particular no meaning
about geometric and physical concepts. Hence, if our world is
invariant under {\it left} $\leftrightarrow$ {\it right}, then
mere information is unable to distinguish between {\it left} and
{\it right}. Now, information is physical, as Landauer used to
emphasize and as every physicists knows today, hence let's
consider classical bits physically realized in some system. For
example the bits 0 and 1 could be realized by right-handed and
left-handed gloves, respectively. It is obvious that such physical
bits can be used to send chirality information. But bits realized
by black and white balls couldn't do the job.

Let us now consider what quantum information brings to the
situation under investigation. Note that recently many authors did
consider similar formal problems, like aligning reference frames,
without paying too much attention to the deep physical meaning of
the problem. At first sight one may quickly conclude that sending
abstract superpositions of basic kets $\ket{0}$ and $\ket{1}$,
i.e. qubits $c_0\ket{0}+c_1\ket{1}$ with complex amplitudes $c_j$,
doesn't help for the same reason as classical bits: they have no
geometric nor physical meaning. The quantum situation is however
more tricky because of the phenomenon of entanglement: Alice and
Bob (we adopt the by now traditional labelling of the two
partners) may exchange many qubits in such a way as to share many
entangled qubit pairs in the singlet state: $\ket{0,1}-\ket{1,0}$.
For clarity assume first that Alice and Bob both use the same
reference frame. Then, using local operations and classical
communication (LOCC), they can measure all the correlations
$\langle\sigma_j\otimes \opone\rangle$,
$\langle\opone\otimes\sigma_k\rangle$ and
$\langle\sigma_j\otimes\sigma_k\rangle$, $j,k=x,y,z$ where
$\sigma_x$, $\sigma_y$ and $\sigma_z$ denote the 3 Pauli matrices.
From these measured correlations Alice and Bob compute the matrix:
\beq \label{matrix} \frac{1}{4}\big(\opone
+\langle\vec\sigma\otimes\opone\rangle\vec\sigma\otimes\opone
+\langle\opone\otimes\vec\sigma\rangle\opone\otimes\vec\sigma
+\langle\sigma_j\otimes\sigma_k\rangle\sigma_j\otimes\sigma_k\big)
\eeq which is nothing but the density matrix representing the
quantum state they share, the projector onto the singlet state in
our example. Consequently, the 4 eigenvalues of (\ref{matrix}) are
positive ($\ge0$). Next, if Bob's reference frame is rotated
compared to Alice's one, but still of the same chirality, then the
matrix (\ref{matrix}) represents the singlet state with one part
accordingly rotated by a unitary transformation; the 4 eigenvalues
would still be positive. What now if Bob's axes are all opposite
to Alice's ones: $\vec r_{Bob}=-\vec r_{Alice}$? This corresponds
to Bob using a frame with opposite chirality with respect to that
of Alice. In such a case it might seem that if Alice and Bob
proceed exactly as above, they would end up with the following
computed matrix: \beq \label{matrixpp} \frac{1}{4}\big(\opone
+\langle\vec\sigma\otimes\opone\rangle\vec\sigma\otimes\opone
-\langle\opone\otimes\vec\sigma\rangle\opone\otimes\vec\sigma
-\langle\sigma_j\otimes\sigma_k\rangle\sigma_j\otimes\sigma_k\big)
\eeq (notice the 2 sign changes!). This amounts to change
$\langle\vec\sigma\rangle$ to $-\langle\vec\sigma\rangle$ on Bob's
side which is equivalent, up to a unitary transformation, to the
well known partial transposition introduced by Peres
\cite{Peres96} as a criteria for entanglement. And indeed the
matrix (\ref{matrixpp}) has a negative eigenvalue whenever the
original state (\ref{matrix}) is entangled \cite{Horodecki96}.
This observation led Lajos Diosi to the initial conclusion that
pure quantum information and entanglement allows one to compare
chiralities. But, as Diosi himself noticed soon after posting
\cite{Diosi00}, this is not quite so. Indeed, the Poincar\'e
sphere that we did implicitly use above does not float in our
"real" 3-dimensional space, but is a convenient mathematical
construction. Abstract qubits don't point in any direction and
thus can't be used to define spatial directions nor chirality.

Let us now assume that the qubits are physically realized by spins
$\half$. One may think that now we have spatial directions and can
thus use entanglement to define chirality. But this is still not
quite so \cite{Rudolph99,CollinsPopescu04}. Indeed, consider a
spin $\half$ particle in a typical Stern-Gerlach experiment and
assume it is deflected towards the ceiling. From this elementary
fact one can't conclude that the resulting spin state is "up"; in
fact the relevant direction is that of the gradient of the
magnetic field. If a physicist using a right-handed reference
frame correctly concludes from the above fact that the spin is
"up", then a physicist using a left-handed frame would equally
correctly conclude that the spin is "down", because the latter
physicist "sees" a magnetic field pointing in the opposite
direction than that assigned (by convention!) by the first
physicist. This is not special to quantum physics, the same holds
in classical physics for all axial vectors, like e.g. angular
momentum. Hence, spins are not appropriate realizations of qubit
if the task is to communicate chirality. Rather than an axial
vector one should use a polar vector like the vector joining an
electron in an atom to its nucleus. This can be done using the
orbital angular momenta $Y^l_m$. But before elaborating on this,
let us look for the closest quantum equivalent to the glove
implementation of classical bits discussed in the introduction.

In order to proceed, we borrow from \cite{Qglove} the concept of
{\it quantum gloves}, that is elementary quantum states that
contain nothing but the abstract concept of chirality, let us
introduce the following chirality operator \cite{Qglove}: \beqa
\Chi&=&\frac{1}{2\sqrt{3}}(\sigma_x\otimes\sigma_y\otimes\sigma_z+\sigma_y\otimes\sigma_z\otimes\sigma_x
+\sigma_z\otimes\sigma_x\otimes\sigma_y \nonumber\\
&-&\sigma_x\otimes\sigma_z\otimes\sigma_y-\sigma_z\otimes\sigma_y\otimes\sigma_x
-\sigma_y\otimes\sigma_x\otimes\sigma_z) \eeqa Note that all
positive permutations appear with positive signs, while all
negative permutations have a negative sign. This chirality
operator has 3 eigenvalues: 0 and $\pm1$, hence
$\Chi=\Chi_+-\Chi_-$. If Alice sends the mixed state
$\rho_+=\half\Chi_+$ corresponding to the projector onto the
2-dimensional eigenspace associated to the eigenvalue +1 and if
Bob measures $\Chi$, then Bob obtains the results +1 if and only
if he uses the same chirality as Alice \footnote{Note that both
states $\rho_\pm$ are invariant under global rotation; they could
thus be used for {\it decoherence free} \cite{decohFreeSupspace}
quantum communication.}. But, if the qubits are realized with
axial vectors, like spins, then knowing $\Chi$ is equivalent to
knowing the chirality, hence in such a case Bob can measure $\chi$
only if he does already know Alice's chirality! Accordingly the
states $\rho_\pm$ are not good abstract {\it quantum gloves}, for
this we need polar vectors, i.e. vectors that really point in some
spatial direction. Since we are not interested in any radial
variable, we now on use the spherical harmonics $Y^l_m$. Under
reflection about the origin, $P\vec n=-\vec n$, they transform as
$PY^l_m=(-1)^lY^l_m$. Consider the two following 4-particle (e.g.
3 electrons and one nucleus) rotationally invariant
states \cite{Qglove}: \beqa |S\rangle &=& Y^0_0Y^0_0Y^0_0\\
 |A\rangle&=& \frac{1}{\sqrt{6}}\big(
Y^1_1Y^1_0Y^1_{-1} + Y^1_0Y^1_{-1}Y^1_{1}+Y^1_{-1}Y^1_{1}Y^1_{0}\nonumber\\
&&
-Y^1_{1}Y^1_{-1}Y^1_{0}-Y^1_{-1}Y^1_{0}Y^1_{1}-Y^1_{0}Y^1_{1}Y^1_{-1}\big)
\eeqa Clearly $P\ket{S}=\ket{S}$ and $P\ket{A}=-\ket{A}$. The
following two states can thus be defined as {\it quantum gloves}:
\beq |G^{\pm}\rangle = { |S\rangle \pm |A\rangle \over \sqrt{2}},
\label{Gpm} \eeq they do indeed transform properly: $P|G^+\rangle
= |G^-\rangle$ and $P|G^-\rangle = |G^+\rangle$!

Note that the two states $|S\rangle$ and $|A\rangle$ can be used
to define qubits $c_0|S\rangle+c_1|A\rangle$: each qubit is
realized with 4 particles. Space reflections act on such qubits as
phase-flips:
$P(c_0|S\rangle+c_1|A\rangle)=c_0|S\rangle-c_1|A\rangle$. The
chirality operator $\Chi$ is then invariant under spatial
reflection: $P\otimes P\otimes P\Chi P\otimes P\otimes P=\Chi$.
Hence, using such realized qubits, Bob could measure $\Chi$
without knowing Alice's chirality. But this requires the use of
3x4=12 particles, the states (\ref{Gpm}) are thus much simpler.

 Like classical information, quantum information per
se is not physical. As emphasized by Landauer, information --
including quantum information -- requires a physical
implementation. Depending on the task, some implementations are
advantageous. The advantage can be so large as to render possible
seemingly impossible tasks. In the present context, the
determination of chirality is impossible using some
implementations of classical bits or of quantum bits, like e.g.
optical pulses and spin ½ respectively, but is perfectly possible
both using c-bits or qubits provided appropriate physical
implementations are used, like e.g. gloves and superpositions of
excited atomic states, respectively \footnote{Note that a
physicist not polluted by excessive use of quantum information
would probably come up with a much simpler solution: send a
circularly polarized photon, the momentum provides the polar
vector and the circular polarization the axial vector, a seemingly
minimal set of 2 vectors.}.

In conclusion, quantum information can be thought of independently
of any implementation, similarly to classical information. This
rather trivial remark implies that quantum information can only
achieve tasks which are expressed in pure information theoretical
terms, like cloning and factoring, but can't perform physical
tasks like aligning reference frames
\cite{Rudolph99,CollinsPopescu04} or defining temperature. This
stresses that quantum teleportation is an information concept and
does not permit the teleportation of a physical object, including
its mass and chirality. This underlines that {\it information is
physical}, but {\it physics is more than mere information}
\cite{infoVsPhysics}.

{\bf Acknowledgements:} This work has been elaborated thanks to --
and much influenced by -- numerous discussions with many
colleagues. In particular, I enjoyed stimulating debates with
Lajos Diosi, Sandu Popescu, Serge Massar and with participants at
the Perimeter Institute's workshop last July on {\it Quantum
Reference Frames}. Financial supports by the swiss NCCR Quantum
Photonics and from the European project RESQ IST-2001-37559 are
gratefully acknowledged.

\end{document}